# Neutronics Analysis of the ITER In-Vessel Viewing System

Andrew Turner[1], Raul Pampin[2], Adrian Puiu[2]

[1]EURATOM/CCFE Fusion Association, Culham Science Centre, Abingdon, Oxon, OX14 3DB, UK
[2] F4E Fusion for Energy, Josep Pla 2, Torres Diagonal Litoral B3, 08019 Barcelona, Spain

*Corresponding author: andrew.turner@ccfe.ac.uk*

## INTRODUCTION

The In-Vessel Viewing System (IVVS) in ITER consists of six identical units which are deployed between pulses or during shutdown, to perform visual examination and metrology of plasma facing components. The system is housed in dedicated ports at B1 level, with deployment at the level between the divertor cassettes and the lowermost outboard blanket modules.

Boron carbide shielding blocks are envisaged to protect the sensitive components of the IVVS from damage during operations, and personnel from radiation fields. In order to progress the design of the IVVS beyond the pre-conceptual stage, analyses were conducted using MCNP to determine the acceptability of a series of different shielding configurations.

The neutron response of the system was assessed using a combination of global variance reduction techniques and a surface source. The effect on radiation loads to neighbouring systems, and various neutronic quantities in the sensitive piezoelectric motors were determined. In addition full 3-D activation and shutdown dose rate calculations were undertaken using MCR2S.

## DESCRIPTION OF THE MODELLING AND ANALYSES

### Model preparation

A CAD model of the IVVS (Fig. 1) was simplified and decomposed using SpaceClaim® and converted into MCNP geometry using the MCAM package [1]. The converted model was integrated into the ITER reference model B-lite v2 (Fig. 2), which was subject to the following modifications:
- B-lite v2 was split and rotated to account for the location of IVVS port plug between lower ports;
- MCNP representations of the neighbouring systems were included: cryopump in the lower port and diagnostics in the equatorial port immediately above the IVVS.
- A realistic bioshield plug was created.
- A new bounding universe for the IVVS and a realistic vacuum vessel penetration were created.

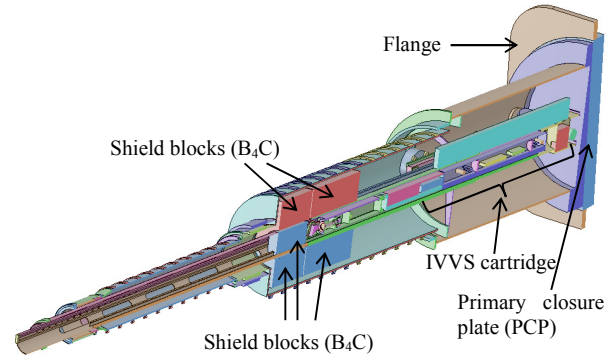

Fig. 1. Cut-away model of the IVVS port and cartridge.

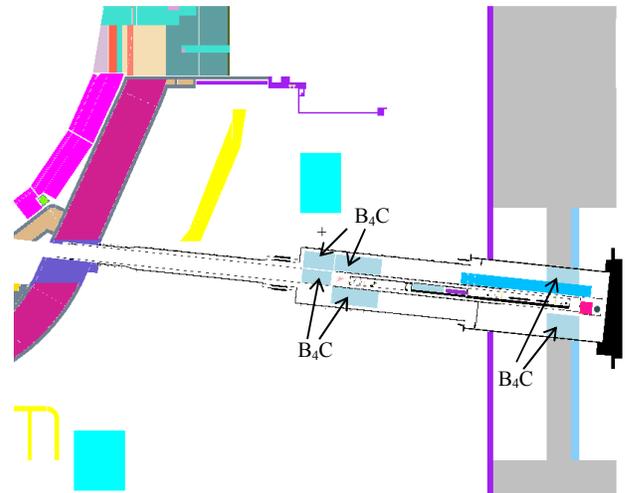

Fig. 2. IVVS (scenario S2) integrated in B-lite v2 showing boron carbide shield blocks.

### Scenarios

Several shielding configurations were modelled, namely:
- S0: the empty IVVS port, with no internal components or shielding, to provide 'baseline' results.
- S1: the IVVS port and internal components, plus boron carbide shield blocks protecting the IVVS probe area.
- S2: as S1, with the addition of boron carbide shield blocks in the vicinity of the bioshield, to provide reduced neutron flux and activation at

- the closure flange and port cell area where dose restrictions apply.
- S2*: as S2, with the removal of the on-axis sliding boron carbide block, to assess the impact of a scenario where the block had failed to return to the correct shielding position.

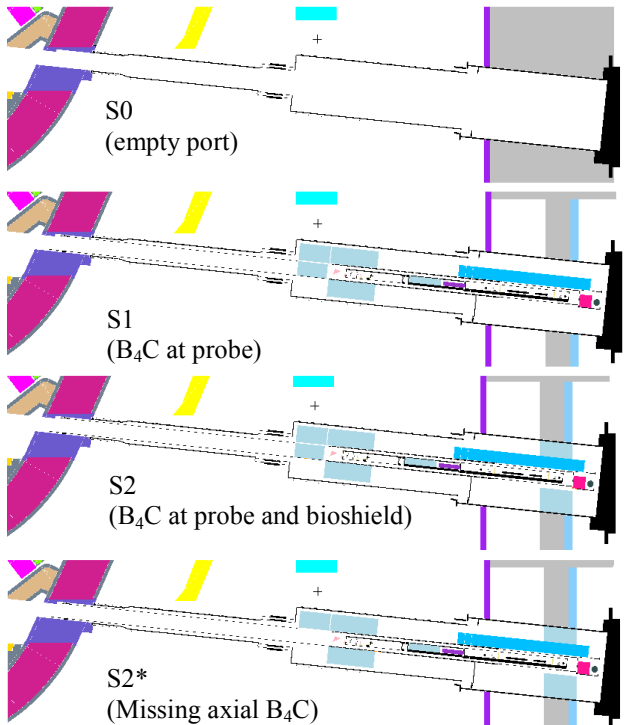

Fig. 3. IVVS shield block configurations S0, S1, S2, S2*

**Methodology**

The neutron response of the system was assessed using MCNP. Global variance reduction (GVR) techniques [2] were required in order to adequately sample the neutrons penetrating the vacuum vessel and equatorial port contributing to the 'global' neutron environment in the IVVS region. It was found that far away from the vessel penetration, the effect of neutrons streaming down the (relatively small) IVVS penetration was comparable to that of the 'global' neutrons; it was also evident that under heavy GVR, the streaming component was not being sufficiently sampled.

To address this, the calculation was performed in two stages: a surface source was produced for neutrons entering the IVVS vessel penetration, which allowed the streaming component to be sampled orders of magnitude more effectively than using the 3-D plasma source; a separate MCNP run was then performed to obtain the 'global' neutron contribution, and in order to prevent double counting, any neutrons passing through the IVVS vessel penetration were 'killed' in this calculation.

Results were obtained for neutron fluxes in the primary closure plate (PCP); the neutron activation of this component having a strong impact on the shutdown dose rate at the port cell area. The effect in radiation loads to neighbouring systems such as divertor pipes, vacuum vessel and magnetic coils was also determined. Additional results were obtained for the absorbed dose rate, helium production and material damage rates in the sensitive piezoelectric motors. For the S2 configuration, coupled neutron-photon analysis was performed to determine nuclear heating effects, and a full 3-D activation calculation undertaken using CCFE's mesh-coupled rigorous two-step code, MCR2S [3]. MCR2S performs an activation calculation under each voxel using the FISPACT-2007 activation code, and produces a shutdown photon source for selected decay times. The activation of large regions of the ITER device were simulated, however due to memory constraints, the activation source was generated on a 6 cm resolution mesh local to the IVVS, and at a 15 cm resolution in neighbouring regions. The resulting photon source was used to determine the shutdown dose rate field from the IVVS removable cartridge (for maintenance considerations) and the absorbed dose rate to the IVVS components due to activation of the IVVS and the surrounding structures in the ITER device.

**RESULTS AND DISCUSSION**

**Shielding analysis results**

The neutron fluxes at the primary closure plate (PCP) are given in Table I and for the piezoelectric (PZT) motors in Table II. Some values were not calculated within MCNP, these estimated results are indicated as (est). Contact dose, damage and helium production rates were determined using FISPACT-II, and thus no statistical errors are available. Table III shows material damage rates and helium production in the PZT actuators.

Table I. PCP neutron flux and contact dose rate

| Scenario | Neutron flux ($n/cm^2/s$) | Contact dose rate at 24 hr ($\mu Sv/hr$) |
|---|---|---|
| S0 | $8.8 \times 10^9$ (6%) | 14,000 |
| S1 | $1.5 \times 10^7$ (3%) | 19.6 |
| S2 | $5.2 \times 10^5$ (9%) | 4.6 |
| S2* | $2.9 \times 10^6$ (30%) | 150 (est) |

Table II. PZT flux and absorbed dose

| Scenario | Neutron flux (n/cm$^2$/s) | EoL absorbed dose (Gy) |
|---|---|---|
| S0 | 2x10$^{10}$ (est) | 2x10$^6$ (est) |
| S1 | 4.3x10$^8$ (20%) | 5x10$^4$ (est) |
| S2 | 3.4x10$^8$ (15%) | 4.07x10$^4$ (11%) |
| S2* | 1.4x10$^{10}$ (8%) | 2x10$^6$ (est) |

Table III. PZT damage and helium production

| Scenario | EoL damage (DPA) | EoL helium (appm) |
|---|---|---|
| S0 | N/C | N/C |
| S1 | 7.3x10$^{-6}$ | 9.5x10$^{-6}$ |
| S2 | 4.7x10$^{-6}$ | 9.4x10$^{-6}$ |
| S2* | 2.0x10$^{-4}$ | 4.1x10$^{-4}$ |

*est=estimated, N/C=not computed, EoL = end of life*

Results for the unshielded configuration, S0, revealed that streaming through the vessel penetration dominates the nuclear response along the IVVS port plug, as expected. Piezoceramic materials have been proven stable up to doses of 1.5x10$^6$ Gy, and the unshielded end-of-life (EoL) absorbed dose to PZT is only marginally below these levels, thus shielding is necessary. In any case, the PCP activation is prohibitively high, since port cell project requirements dictate a limit of 10 μSv/hr after 24 hours of decay time in this area.

The S1 scenario included shielding blocks at probe level, and results showed significantly reduced PZT absorbed dose, however the PCP contact dose rate still marginally exceeded dose limits. In order to further reduce dose levels, additional shielding blocks were modelled inside the port at the level of the bioshield (configuration S2), the results of which showed PCP activation levels below the port cell limit. The S2* configuration showed, again, unacceptable levels of PCP activation.

**S2 full analysis results**

Results of the more comprehensive analysis conducted for scenario S2 provided additional data. The nuclear heating during plasma operation in the axial shield block is ~1 W, ~80% of which is deposited in the first 10 cm; very low heating was found in other shield blocks and IVVS components. During shutdown, the PZT absorbed dose and heating rates were found to be several orders of magnitude lower than those during a plasma pulse, and not reported here. Table IV gives the shutdown dose rate in the vicinity of an isolated IVVS cartridge.

Table IV. Shutdown dose rate (μSv/hr) for scenario S2

| Distance | Dose rate (μSv/hr) | | |
|---|---|---|---|
| | Time 10$^4$ s | 10$^5$ s | 10$^6$ s |
| Contact | 580 | 160 | 140 |
| 20 cm | 170 | 43 | 39 |
| 100 cm | 35 | 8.8 | 8.0 |

The shutdown dose rate in contact with the isolated IVVS cartridge was found to exceed the hands-on maintenance limit of 100 μSv/hr. A rigorous treatment of uncertainty propagation is very challenging and a topic of current research, however experience suggests a good estimate of the final uncertainty on dose rate is up to twice that of the total neutron flux uncertainty. The total neutron fluxes had errors under 10% on a relatively fine (6 cm) mesh, suggesting a high degree of confidence in the results and an estimated uncertainty of 20-30%.

The peak neutron flux, material damage rate and helium production rate were assessed for the divertor cooling pipes and vacuum vessel in the region immediately behind the cut-out in the blanket and triangular support (Table V).

Table V. Effect of the IVVS cut-out

| Location | Flux (n/cm$^2$/s) | EoL damage (DPA) | EoL Helium (appm) |
|---|---|---|---|
| Divertor pipe | 3.17x10$^{12}$ | 1.30x10$^{-2}$ | 2.20x10$^{-1}$ |
| Vacuum vessel | 3.11 x10$^{12}$ | 9.22x10$^{-3}$ | 1.92x10$^{-1}$ |

EoL = end of life

The presence of the cut-out was found to increase damage rates and helium production in the divertor pipe region and vacuum vessel by a factor of 3 immediately behind the cut-out, when compared with values obtained for a calculation with no such cut-out. The values are still within ITER project limits. Additionally it was demonstrated that the IVVS vessel penetration has a negligible effect on magnetic coil nuclear loads.

**CONCLUSIONS**

Analyses were conducted using MCNP to determine the acceptability of the current IVVS design for a series of shielding configurations. Four different geometry scenarios were considered. Results were obtained for the neutron flux, nuclear heating and absorbed dose rates, helium production and material damage rates, as well as for the neutron flux and spectrum in the closure plate. The shutdown dose rate was assessed for scenario S2, both for activation of the IVVS and surrounding ITER components, and around the removable IVVS cartridge in isolation. The effect of the IVVS penetration in radiation loads to neighbouring systems such as divertor pipes, vacuum vessel and magnetic coils was determined.

Analysis on the pre-conceptual design has concluded that:

- In an unshielded configuration, the end-of-life (EoL) absorbed dose to PZT is only marginally below stable limits. Activation of the PCP is prohibitively high in such a configuration.
- S1 scenario, with shielding blocks at probe level, showed significantly reduced PZT absorbed dose, however the PCP contact dose rate exceeded dose limits (though only by a factor of 2).
- The presence of additional shielding blocks at the level of the bioshield (configuration S2), demonstrated PCP activation levels below the port cell limit.
- The S2* configuration, in which the movable shield block failed to return to the correct position, showed unacceptable levels of PCP activation.
- Absorbed dose and heating rates at shutdown to the PZT are several orders of magnitude lower than those during a plasma pulse and are not significant.
- The effect of the IVVS cut-out in the blanket and triangular support is to increase damage rates and helium production in the immediate vicinity by a factor of ~3. The absolute values for the vacuum vessel and divertor cooling pipes are still within project limits.
- The IVVS vessel penetration has a negligible effect on magnetic coil nuclear loads.
- Nuclear heating rates in the shielding blocks and other IVVS components are very small. The nuclear heating rate in the axial shield block is ~1 W.
- Shutdown dose rates in contact with the isolated IVVS cartridge (a maintenance scenario) are in the range 100-160 μSv/hr at $t=10^5$ seconds, only marginally lower at $t=10^6$ seconds and exceeding hands-on maintenance limits of 100 μSv/hr.

The modelling and calculations reported here assisted the pre-conceptual engineering design of the IVVS for ITER. The surface source approach greatly improved analysis turnover and statistics, and provided valuable information on the relative magnitude of these components for the different scenarios analysed. The IVVS design is now progressing and further analyses will be required to guarantee consistency and performance.

The S1 scenario is preferred over S2 for engineering feasibility, and these analyses have shown PCP contact dose for this the S1 scenario is only marginally above requirements, and requires further consideration in the next stage of the design process. In addition, alternative low-activation materials for the cartridge are being considered, and more detailed modelling of the radiation fields in the port cell area is being performed.

*This work was carried out using an adaptation of the B-lite MCNP model which was developed as a collaborative effort between the FDS team of ASIPP China, University of Wisconsin-Madison, ENEA Frascati, CCFE UK, JAEA Naka, and the ITER Organization.*

*This work was funded by F4E under contract F4E-2008-OPE-002-01-06. The views and opinions expressed herein do not necessarily reflect those of F4E or the ITER Organization. F4E is not liable for any use that may be made of the information contained herein.*